\documentclass[npg, manuscript, online]{copernicus}

\begin{document}
\nolinenumbers

\title{Parametrization of stochastic multiscale triads}

\Author[1,2]{Jeroen}{Wouters}
\Author[3]{Stamen I.}{Dolaptchiev}
\Author[2,4,5]{Valerio}{Lucarini}
\Author[3]{Ulrich}{Achatz}

\affil[1]{School of Mathematics and Statistics, The University of Sydney, Sydney, Australia}
\affil[2]{Klimacampus, Meteorologisches Institut, University of Hamburg, Hamburg, Germany}
\affil[3]{Institut f{\"u}r Atmosph{\"a}re und Umwelt, Goethe-Universit{\"a}t Frankfurt, Frankfurt am Main, Germany}
\affil[4]{Department of Mathematics and Statistics, University of Reading, Reading, UK}
\affil[5]{Walker Institute for Climate System Research, University of Reading, Reading, UK}

\runningtitle{Parametrization of stochastic multiscale triads}

\runningauthor{Wouters, Dolaptchiev, Lucarini, Achatz}

\correspondence{Jeroen Wouters (jeroen.wouters@uni-hamburg.de)}

\received{}
\pubdiscuss{} 
\revised{}
\accepted{}
\published{}

\firstpage{1}

\maketitle

\begin{abstract}
  We discuss applications of a recently developed method for model reduction
  based on linear response theory of weakly coupled dynamical systems. We
  apply the weak coupling method to simple stochastic differential equations
  with slow and fast degrees of freedom. The weak coupling model reduction
  method results in general in a non-Markovian system, we therefore discuss
  the Markovianization of the system to allow for straightforward numerical
  integration. We compare the applied method to the equations obtained through
  homogenization in the limit of large time scale separation between slow and
  fast degrees of freedom. We numerically compare the ensemble spread from a fixed initial
  condition, correlation functions and exit times from domain. The weak
  coupling method gives more accurate results in all test cases, albeit with a higher
  numerical cost.
\end{abstract}

\introduction
Many models of physical systems are too complex to be solved analytically, or
even numerically if a large range of temporal and spatial scales is involved.
For some high-dimensional dynamical systems it is however possible to derive
lower-dimensional reduced models
{\citep{givon_extracting_2004,huisinga_extracting_2003}}. The reduced model is
easier to solve analytically and faster to integrate numerically, while still
preserving some of the essential characteristics of the full system. This line
of research lies at the heart of many applications, for example in molecular
dynamics {\citep{hijon_morizwanzig_2009,lu_exact_2014}} and climate modeling
{\citep{lucarini_mathematical_2014,imkeller_stochastic_2001,palmer_stochastic_2009}}.

The derivation of a reduced model is possible, for example, in the presence of a
time scale separation between slow resolved and fast unresolved variables, as is
assumed in the homogenization method {\citep{pavliotis_multiscale_2008}}. This
method applies to slow-fast systems of the form
\begin{eqnarray}
  \dot{x} & = & f_0 (x, y) + \frac{1}{\varepsilon} f_1 (x, y)\nonumber\\
  \dot{y} & = & \frac{1}{\varepsilon^2} g_1 (x, y) + \frac{1}{\varepsilon^{}}
  \beta (y) \xi (t), \label{eq.slowfast}
\end{eqnarray}
in the limit of infinite time scale separation $\varepsilon \rightarrow 0$,
where $\xi$ denotes a standard Brownian motion (i.e. the equations should be
considered equivalent to a stochastic integral in the It{\^o} interpretation) \citep{khasminskii_principle_1963,papanicolaou_probabilistic_1976}.
It is evident from the dynamical equation that the $y$ variables evolve on a
faster time scale than the $x$ variables. For finite values of $\varepsilon$
there is an intricate feedback between the evolution of the $x$ and $y$
variables. The situation simplifies in the limit of $\varepsilon \rightarrow
0$ where the slow variables do not evolve on the time scales on which $y$
strongly fluctuates. As a result, the slow dynamics converges to a stochastic
evolution, where the effect of $y$ is completely replaced by statistical
quantities related to the motion of $y$ for a fixed value of $x$. On a more
technical note, the precise expression for the quantities entering in the
reduced dynamics can be easily obtained through an expansion in $\varepsilon$
of the backward Kolmogorov equation (the adjoint of the Fokker-Planck
equation) $\partial_t v (x, t) = (\mathcal{L}_0 +\mathcal{L}_1 / \varepsilon
+\mathcal{L}_2 / \varepsilon^2) v (x, t)$ of corresponding to the slow-fast
dynamics (where $\mathcal{L}_0 = f_0 \partial_x$, $\mathcal{L}_1 = f_1
\partial_x$ and $\mathcal{L}_2 = g_1 \partial_y + (\beta/2) \partial_y^2)$ \citep{pavliotis_multiscale_2008}.

The method of homogenization has found a great number of applications in
different fields of physics and mathematics
{\citep{pavliotis_multiscale_2008}}. Many physical systems, however, do not
feature a time scale separation. As an example, the climate system has variability on many different temporal (and spatial) scales, but no clear spectral gaps can be identified. This creates fundamental difficulties in the theoretical investigation of climate dynamics and in the construction of climate models. As a result, approximate equations are used for dealing with scales of motions belonging to a range of scales of interest, and numerical models are able to resolve explicitly only a fractions of the full range of scales. The dynamics taking place on scales that are too small and/or fast to be resolved need to be parametrized. Consider the case of
convective motion in the Earth's atmosphere. Convective clouds are significant for the climate, yet can
only be resolved at a spatial resolution of 10--100 m
{\citep{sakradzija_fluctuations_2015}}, whereas climate models only resolve
scales of the order of 100 km
{\citep{intergovernmentalpanelonclimatechange_climate_2013}}. Unresolved
convective motion however features a so-called “gray zone”, a range of time scales
overlapping with the dynamical time scales of the resolved large scale flow
{\citep{sakradzija_fluctuations_2015}}, therefore homogenization can not be
applied. It is a formidable challenge to derive dimension reduction methods
that do not require a time scale separation. One should underline that when facing a lack of time scale separation, we would like to be able to construct self-adaptive parametrizations as opposed to empirical ones, so that when the resolution of a numerical model is changed we do not need to redo the exercise of fitting a reduced model.

Going beyond the familiar setting of infinite time scale separation requires a
novel approach to the derivation of closed equation for the reduced system.
Recently, we have developed a model reduction technique that does not rely on
the presence of such a separation
{\citep{lucarini_mathematical_2014,wouters_disentangling_2012,wouters_multilevel_2013}}.
The alternative method for model reduction makes use of a weak coupling
approach, in which response theory
{\citep{ruelle_review_2009,ruelle_differentiation_1997}} is used to derive a
closure. The systems of interest follow a dynamics determined by
\begin{eqnarray}
  \dot{x} & = & \varepsilon \psi_x (x, y) + f_x (x)\nonumber\\
  \dot{y} & = & \varepsilon \psi_y (x, y) + g_y (y),\label{eq.weaklycoupled}
\end{eqnarray}
where $x$ is the variable of interest. Exploiting the weak coupling form of
this equation, response theory can be employed to expand expectation values of
$x$-dependent observable under the invariant measure in orders of $\varepsilon$. This expansion yields a
series in terms of $\varepsilon$, reminiscent of the Dyson series in
scattering theory, each representing a sequence of interactions between the
$x$ and $y$ subsystems, corresponding to a certain Feynman diagram.

The truncation of this series up to a given order yields an approximation of
the response of the $x$ subsystem to the coupling to the $y$ subsystem. More
importantly, it allows to determine the statistical quantities of the $y$
system that dictate this response. The first order correction to the dynamics
of the $x$ system can be written as the expectation value $\varepsilon \int \mathrm{d}
y \psi_x (x, y) \rho_y (y)$, where $\rho_y$ is the invariant density of the
uncoupled $\dot{y} = g_y (y)$ dynamics. At second order two
correction terms appear, one due to double $\psi_x$ interactions from $y$ to $x$,
determined by a correlation function of the uncoupled $y$ dynamics, and a feedback
term, determined by a response function of the uncoupled $y$ dynamics. This knowledge can then be exploited to
derive a surrogate dynamics for $x$ that reproduces the effect of the coupling
of $x$ to $y$ up to second order in $\varepsilon$.
While this theory has been originally developed assuming that the uncoupled
systems are Axiom A dynamical systems, it can be equally applied in the case
where the uncoupled dynamics is stochastic, the only needed requirement being
to have a physical measure.
Interestingly, the results obtained using response theory match what one can
derive by constructing a perturbative expansion of the dynamics of the system
using the Mori-Zwanzig projection method {\citep{wouters_multilevel_2013}}.

Previously, we have proposed a surrogate dynamical equation for the $x$
variable that introduces an $\varepsilon$-dependent perturbing term to the
dynamics $f_x$ to match the response of the statistics of the full system. The
perturbing term contains a non-Markovian memory term and a correlated noise,
with the memory kernel and correlation functions depending on the statistics
of the uncoupled dynamics $\dot{y} = g_y$. In a recent study of the applicability of the
weak coupling approach to a simple ocean-atmosphere system, the method has
been shown to give a good result for sufficiently weak coupling between the
ocean and the atmosphere {\citep{demaeyer_stochastic_2016}}, even if it is clear that a systematic investigation of the performance of the weak coupling approach is indeed still needed.

We remark that Chekroun et al.
{\citep{chekroun_stochastic_2015,chekroun_approximation_2015}} have recently
proved that, indeed, constructing reduced order models entails introducing
deterministic, stochastic and memory correction to the dynamics of the
variables of interest.

Here we will apply and extend the weak coupling approach
  of \citep{wouters_disentangling_2012,wouters_multilevel_2013} for the development of parameterizations for various
  stochastic triad models. Triad interactions arise from quadratic
  nonlinearities with energy conserving properties (see e.g., \citep{gluhovsky_structure_1999}).
  The triad models considered here appear in applications of
  the homogenization technique
  to construction of parameterizations in climate
  modeling (see e.g., \citep{MTV01, MTV02, FMV05, FM06, dotiac, doacti}). The
non-Markovian memory kernel in the weak coupling approach will be calculated for these simple stochastic multiscale models and
approximated by a Markovian stochastic process, in order to allow for
easier numerical implementation. The systems we investigate can be written in
both the weak coupling form of Eq. \ref{eq.weaklycoupled} and the slow-fast form of Eq. \ref{eq.slowfast}, therefore direct
comparison is possible and will be performed on a number of metrics, namely initial
ensemble spread, correlation functions and exit times from an interval.

\section{The additive triad}

The first model we look at is the stochastically forced additive triad.
This system is a low-dimensional model that has non-linear interactions
reminiscent of those occurring between the Fourier modes of a fluid flow. It is
stochastically forced to mimic the interaction with further unresolved modes.
The system has three variables, one slow variable $x$ and two fast variables
$y_1$ and $y_2$. The fast dynamics is dominated by two independent
Ornstein-Uhlenbeck processes. The dynamical equations for this triad are
\begin{eqnarray}
  \frac{d x}{d t} & = & B^{(0)} y_1 y_2  \nonumber\\
  \frac{d y_1}{d t} & = & B^{(1)} xy_2 - \frac{\gamma_1}{\varepsilon} y_1 +
  \frac{\sigma_1}{\sqrt{\varepsilon}} \xi_1 (t) \nonumber\\
  \frac{d y_2}{d t} & = & B^{(2)} xy_1 - \frac{\gamma_2}{\varepsilon} y_2 +
  \frac{\sigma_2}{\sqrt{\varepsilon}} \xi_2 (t) \,.  \label{eq.triad}
\end{eqnarray}
The processes $\xi_i$ are independent Brownian motions in the It{\^o} sense.
Here and below a differential equation featuring a Brownian motion will be
interpreted as the equivalent stochastic integral. In addition, we require $\sum_i B^{(i)}
  = 0$, which guarantees energy conservation in the case
  $\gamma_i=\sigma_i=0$.

\subsection{Homogenization}

On the time scale $t$, when increasing the time scale separation $1 /
\varepsilon$ to infinity, we have trivial dynamics of the averaged equations
$\dot{\bar{x}} = B^{(0)} \langle y_1 y_2 \rangle_{\rho_{\text{OU}}} = 0$ where
$\rho_{\text{OU}}$ is the Gaussian invariant measure of the fast
Ornstein-Uhlenbeck process generated by taking $B^{(i)} = 0$ for $i = 1, 2,
3$. In the setting of homogenization, one looks at the convergence of the
distribution of paths on a longer time scale. The time is scaled to the
diffusive time scale $\theta = \varepsilon t$ and on this longer diffusive time
scales deviations from the averaged dynamics develop.

\begin{figure}[t]
\includegraphics[width=8.3cm]{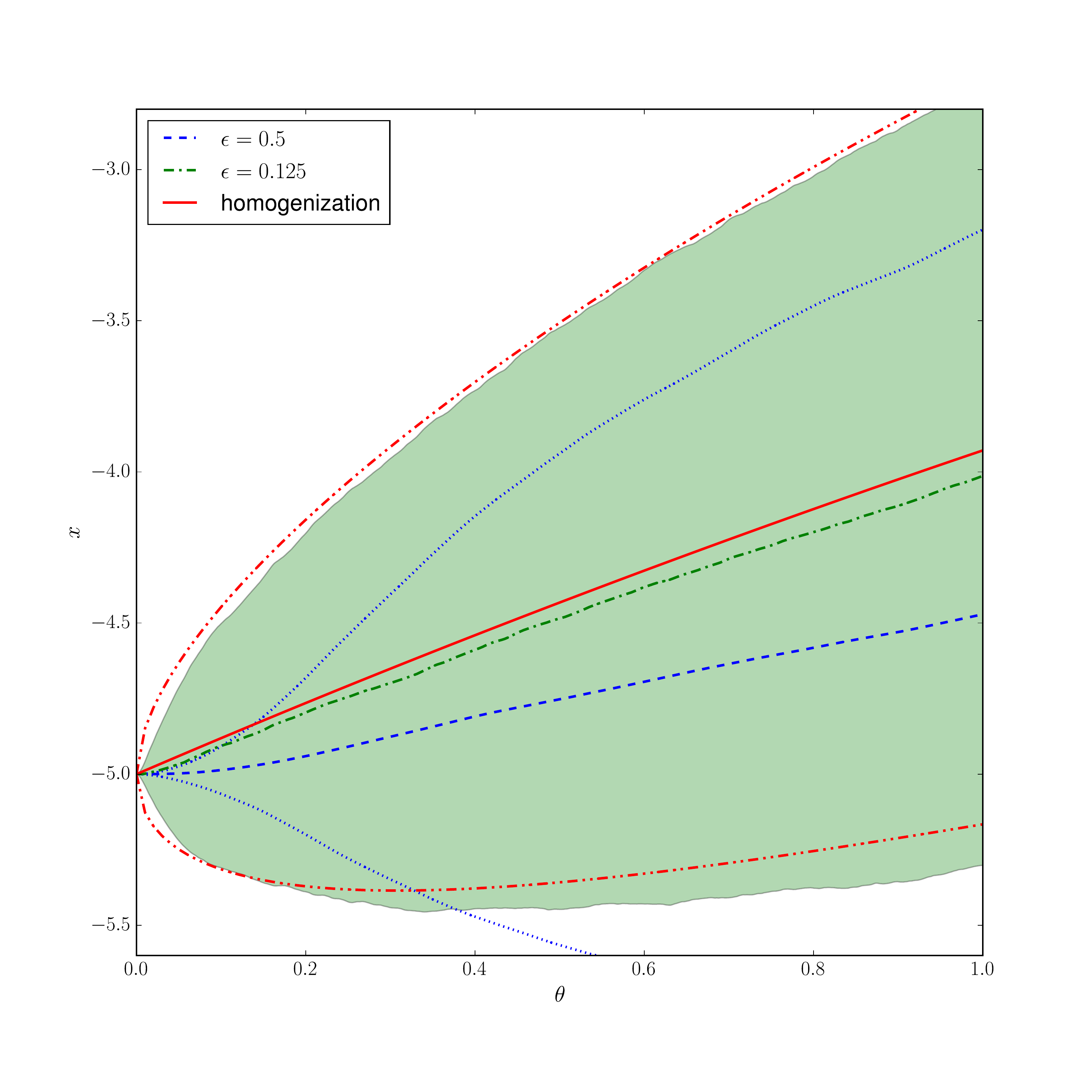}
\caption{Convergence to the homogenized equations for the additive triad
  (\ref{eq.triad}) in $\theta = \varepsilon t$ time scale. The red solid and
  dash-double-dotted lines show the mean and $2 \sigma$ intervals respectively
  for an ensemble evolving according to the homogenized equation
  (\ref{eq.addHom}) from an initial condition $x = - 5$. The blue dashed and
  dotted lines show the mean and $2 \sigma$ intervals for an ensemble of the
  additive triad (\ref{eq.triad}) for $\varepsilon = 0.5$ from an initial
  condition $(x, y_1, y_2) = (- 5, 0, 0)$ with $B^{(0)} = - 0.75$, $B^{(1)} =
  - 0.25$, $B^{(2)} = 1$, $\gamma_1 = 1 / \delta$, $\sigma_1 = \sqrt{2 /
  \delta}$, $\gamma_2 = 1$ and $\sigma_2 = \sqrt{2}$ with $\delta = 0.75$. The
  green dash-dotted line and the green shaded area show the same for
  $\varepsilon = 0.125$. \label{fig.addHom}}
\end{figure}

By expanding the backward Kolmogorov equation for the slow-fast system in
orders of $\varepsilon$, a Kolmogorov equation for only the slow variables can
be derived (see {\citep{pavliotis_multiscale_2008}}). The dynamical equation
corresponding to this Kolmogorov equation is in this case a one-dimensional
Ornstein-Uhlenbeck process \citep{MTV02}
\begin{eqnarray}
  \frac{\partial x}{\partial \theta} & = & C_m x + \sqrt{2 A_0} \xi (\theta) \,,  \label{eq.addHom}
\end{eqnarray}
where
\begin{eqnarray*}
  C_m & = & \frac{B^{(0)}}{\gamma_1 + \gamma_2}  \left( B^{(1)}
  \frac{\sigma_2^2}{2 \gamma_2} + B^{(2)} \frac{\sigma_1^2}{2 \gamma_1}
  \right)\\
  A_0 & = & \frac{{B^{(0)}}^2}{\gamma_1 + \gamma_2}  \frac{\sigma_1^2 }{2
  \gamma_1 } \frac{\sigma_2^2}{2 \gamma_2} \,.
\end{eqnarray*}
See Fig. \ref{fig.addHom} for an illustration of the homogenization
principle for the additive triad. The mean and variance of the triad converge
to those of the Ornstein-Uhlenbeck process (\ref{eq.addHom}) for small
$\varepsilon$.

\subsection{Weak coupling limit}

We will now discuss the weak coupling method as described in
{\citep{wouters_disentangling_2012,wouters_multilevel_2013}}. By rescaling the
time as $\tau = \varepsilon^{-1} t$ we can write the
stochastically forced additive triad equation (\ref{eq.triad}) as a
two-dimensional Ornstein-Uhlenbeck system weakly coupled non-linearly to a
trivial zero-gradient $x$ system:
\begin{eqnarray}
  \frac{d x}{d \tau} & = & \varepsilon \psi_x (y_1, y_2) \nonumber\\
  \frac{d y_1}{d \tau} & = & \varepsilon \psi_{y, 1} (x, y) - \gamma_1 y_1 +
  \sigma_1 \xi_1(\tau) \nonumber\\
  \frac{d y_2}{d \tau} & = & \varepsilon \psi_{y, 2} (x, y) - \gamma_2 y_2 +
  \sigma_2 \xi_2(\tau) \, .  \label{eq.coupled}
\end{eqnarray}
with $\psi_x(y_1, y_2) = B^{(0)} y_1 y_2$ and $\psi_y(x,y) = (B^{(1)} x y_2, B^{(2)} x y_1)^T$.
The stochastic parametrization derived in
{\citep{wouters_disentangling_2012,wouters_multilevel_2013}} is given by a
non-Markovian equation
\begin{eqnarray}
  \frac{d \tilde{x}}{d \tau} & = & \varepsilon \sigma (\tau) + \varepsilon^2
  \int_0^{\infty} \mathrm{d} sR (s, \tilde{x} (\tau - s))\, ,  \label{eq.memoryparam}
\end{eqnarray}
where the the memory kernel $R(s,\tilde x)$ and first two moments of the stochastic process $\sigma(\tau)$
are derived using the weak coupling method to the following statistics of the
uncoupled $y$ Ornstein-Uhlenbeck dynamics:
\begin{eqnarray}
  \langle \sigma (\tau) \rangle & = & 0 \nonumber\\
  C (\tau) := \langle \sigma (0) \sigma (\tau) \rangle & = & \langle
  \psi_x (y_1, y_2) \psi_x (y_1(\tau), y_2(\tau))
  \rangle_{\rho_{OU}}  \label{eq.triadCorr}\\
  R (\tau) & = & \langle \psi_y (x, y_1, y_2) . \nabla_y \psi_x (
  y_1(\tau), y_2(\tau)) \rangle_{\rho_{OU}}\, .  \label{eq.triadMem}
\end{eqnarray}
where the evolution of $y_1$ and $y_2$ into $y_1(\tau)$ and $y_2(\tau)$ are taken to be the uncoupled Ornstein-Uhlenbeck dynamics $dy_i/d \tau = - \gamma_i y_i + \sigma_i \xi_i$.
We have for the case of the additive triad (\ref{eq.triad})

\begin{align}
  C (\tau) = (B^{(0)})^2  \langle y_1 (0) y_1 (\tau) \rangle  \langle y_2
  (0) y_2 (\tau) \rangle_{\rho_{OU}} = (B^{(0)})^2 \exp (- (\gamma_1 +
  \gamma_2) \tau) \frac{\sigma_1^2}{2 \gamma_1}  \frac{\sigma_2^2}{2 \gamma_2} \label{eq.triadCorr2}
\end{align}

and

\begin{align}
  R (\tau, x) = & B^{(0)} B^{(1)} x \langle y_2 (0) (\partial_{y_1} y_1
  (\tau)) y_2 (\tau) \rangle_{\rho_{OU}} \nonumber\\
  & + B^{(0)} B^{(2)} x \langle y_1 (0) y_1 (\tau) (\partial_{y_2} y_2
  (\tau)) \rangle_{\rho_{OU}} \nonumber \\
  = & xB^{(0)} \exp (- (\gamma_1 + \gamma_2) \tau) \left( \frac{\sigma_2^2}{2
  \gamma_2} B^{(1)} + \frac{\sigma_1^2}{2 \gamma_1} B^{(2)} \right) \, . \label{eq.triadMem2}
\end{align}

\subsubsection{Markovian parametrization}

Due to the identical time-scale $\gamma_1 + \gamma_2$ in both memory and noise
correlation, the memory equation (\ref{eq.memoryparam}) can be transformed to
a Markovian parametrization. We want to find a parametrizing two level
Markovian dynamical system of the form
\begin{eqnarray}
  \frac{d z_1}{d \tau} & = & \varepsilon C_1 z_2 \nonumber\\
  \frac{d z_2}{d \tau} & = & - \gamma^{} z_2 + \sigma_z \xi (\tau) + \varepsilon
  C_2 z_1 \, .  \label{eq.memoryparamMarkov}
\end{eqnarray}
such that the second order response of this system to changes in $\varepsilon$ is the same as the response of (\ref{eq.memoryparam}). In other words, we want to determine the parameters $C_1$, $C_2$, $\gamma$ and $\sigma_z$ in (\ref{eq.memoryparamMarkov}) such that the correlation and memory functions of the fast equation in (\ref{eq.memoryparamMarkov}) are equal to (\ref{eq.triadCorr2}) and (\ref{eq.triadMem2}) respectively. The correlation function $C(\tau)$ and memory function $R(\tau)$ of the fast equation of
(\ref{eq.memoryparamMarkov}) are
\begin{eqnarray}
  C (\tau) & = & \langle (C_1 z_2(0)) (C_1 z_2(\tau)) \rangle = C_1^2 e^{- \gamma \tau} \frac{\sigma_z^2}{2 \gamma} \\
  R (\tau, z_1) & = & \langle (C_2 z_1) \partial_{z_2} (C_1 z_2(\tau)) \rangle = C_1 C_2 z_1 e^{- \gamma \tau} \, ,
\end{eqnarray}
where the evolution of $z_2$ to $z_2(\tau)$ is now given by $d z_2 /d \tau = -\gamma z_2 + \sigma_z \xi(\tau)$.
By equating these functions to their counterparts in (\ref{eq.triadCorr2}) and (\ref{eq.triadMem2}) we see that by choosing
\begin{eqnarray*}
  C_1 & = & B^{(0)}\\
  C_2 & = & \frac{\sigma_2^2}{2 \gamma_2} B^{(1)} + \frac{\sigma_1^2}{2
  \gamma_1} B^{(2)} = \beta_2 B^{(1)} + \beta_1 B^{(2)}\\
  \gamma & = & \gamma_1 + \gamma_2\\
  \sigma_z^2 & = & 2 \frac{\sigma_1^2}{2 \gamma_1} \frac{\sigma_2^2}{2
  \gamma_2} (\gamma_1 + \gamma_2) = 2 \beta_1 \beta_2 \gamma
\end{eqnarray*}
the reduced $z_1$ dynamics of the parametrized dynamical system in the weak
coupling method are of the same form as those of the stochastic triad
(\ref{eq.triad}).

This Markovian reduced equation (\ref{eq.memoryparamMarkov}) is in fact a
reformulation of the non-Markovian equation (\ref{eq.memoryparam}). To see
this, we write an explicit solution for $z_2$ in function of the history of
$z_1$ and $\xi$ as
\begin{eqnarray*}
  z_2 (\tau) & = & e^{- \gamma \tau} z_2 (0) + \int_0^{\tau} \mathrm{d} t'
  (\sigma_z \xi (t') + \varepsilon C_2 z_1 (t')) e^{- \gamma (\tau - t')} \, .
\end{eqnarray*}
This solution can then be inserted into (\ref{eq.memoryparamMarkov}), to
obtain
\begin{eqnarray}
  \frac{d z_1}{d \tau} & = & \varepsilon C_1 e^{- \gamma \tau} z_2 (0) +
  \varepsilon C_1 \int_0^{\tau} \mathrm{d} t' (\sigma_z \xi (t') + \varepsilon C_2 z_1 (t'))
  e^{- \gamma (\tau - t')} \, ,
\end{eqnarray}
which agrees with (\ref{eq.memoryparam}), the first two terms being an
Ornstein-Uhlenbeck process with the required correlation plus a memory term
with the required memory kernel.

\begin{figure*}[t]
  \includegraphics[width=12cm]{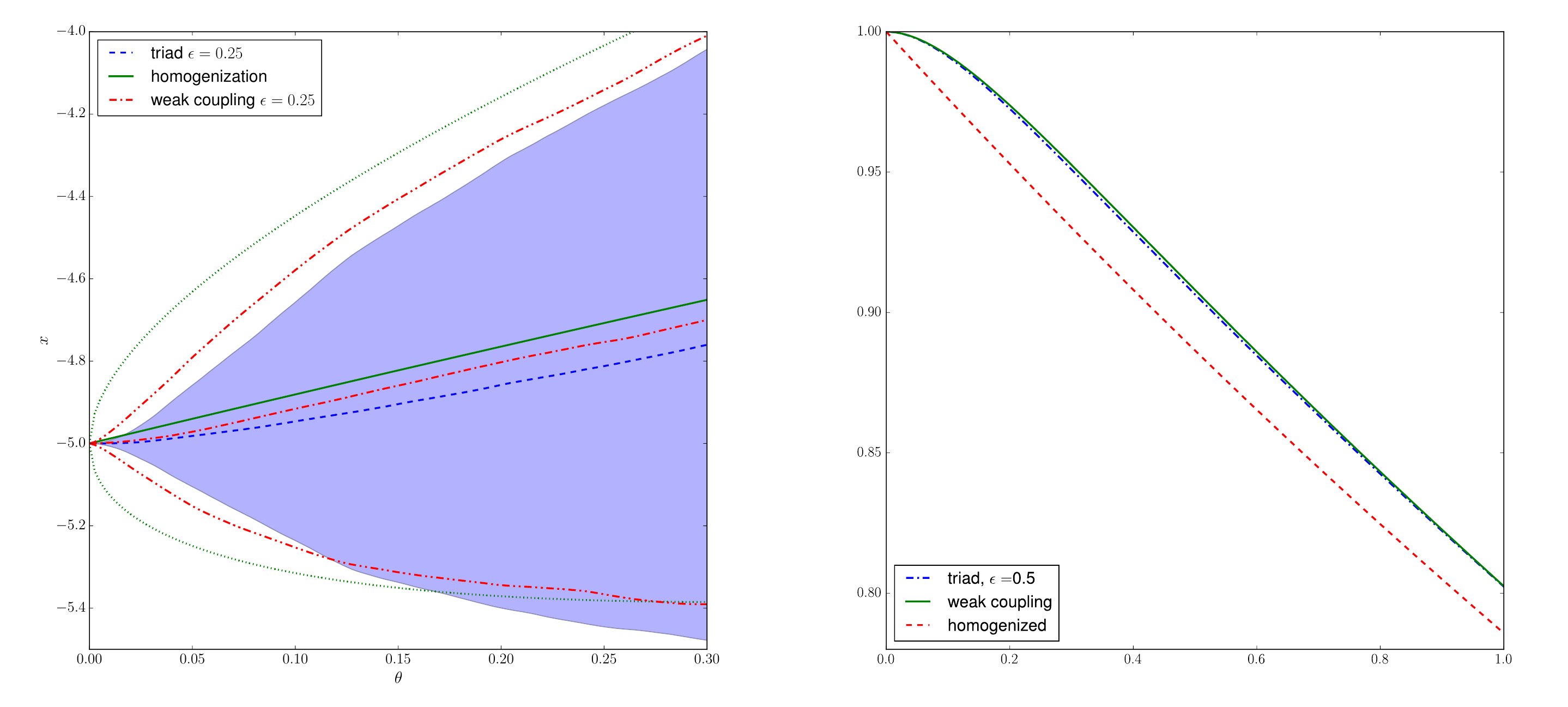}
  \caption{Left: comparison of the ensemble spread for the original additive
  triad system for $\varepsilon = 0.25$ from an initial condition $(-5,0,0)$
  (the ensemble mean is the blue dashed line, $2 \sigma$ interval the blue
  shaded area), the two-level Ornstein-Uhlenbeck process from the weak
  coupling method (\ref{eq.memoryparamMarkov}) from an initial condition $(-5, 0)$
  (ensemble mean: red dash-dotted line, $2 \sigma$ interval: red
  dash-dot-dotted lines) and the one-level Ornstein-Uhlenbeck process from
  homogenization (\ref{eq.addHom}) from $x = - 5$ (ensemble mean: green solid
  line, $2 \sigma$ interval: dotted lines)\\
  Right: comparison of the autocorrelation functions of the slow variable
  $\langle x (t) x (0) \rangle$ in the full triad for $\varepsilon = 0.5$
  (blue dash-dotted line), $\langle z_1 (t) z_1 (0) \rangle$ in the weak
  coupling model (green solid line) and $\langle x (t) x (0) \rangle$ for the
  homogenized equation (red dashed line).\\
  Both plots use parameter values $B^{(0)} = - 0.75$, $B^{(1)} = - 0.25$,
  $B^{(2)} = 1$, $\gamma_1 = 1 / \delta$, $\sigma_1 = \sqrt{2 / \delta}$,
  $\gamma_2 = 1$ and $\sigma_2 = \sqrt{2}$ with $\delta = 0.75$.\label{fig.addCompar}}
\end{figure*}

This Markovian formulation allows for a straightforward numerical
implementation of the parametrization, compared to the non-Markovian equation
(\ref{eq.memoryparam}) which requires one to store the history of the process
in memory.

A comparison of the performance of the two model reductions is show in Figure
\ref{fig.addCompar}. Shown are the spread of an ensemble initiated at a fixed
value for the slow variables $x = z_1 = - 5$ and the autocorrelation function
of the slow variables. The weak coupling method clearly gives better results.

By correctly rescaling time and taking the limit of $\varepsilon \rightarrow 0$
in the Markovian parametrization (\ref{eq.memoryparamMarkov}) one can
furthermore verify that in this limit it converges to the homogenization of
the original triad equation (Eq. (\ref{eq.addHom})).

\section{The slowly oscillating additive triad}

The additive triad as specified in Eq. (\ref{eq.triad}) can be generalized to
allow for an additional interaction between the $y$ variables on the slow time
scale that is independent of $x$. The dynamical equations for this slowly
oscillating triad are
\begin{eqnarray}
  \frac{d x_{}}{d t} & = & B^{(0)} y_1 y_2 \nonumber\\
  \frac{d y_1}{d t} & = & B^{(1)} y_2 x - \frac{\gamma_1}{\varepsilon} y_1 +
  \omega y_2 + \frac{\sigma_1}{\sqrt{\varepsilon}} \xi_1 (t) \nonumber\\
  \frac{d y_2}{dt} & = & B^{(2)} x_{} y_1 - \frac{\gamma_2}{\varepsilon}
  y_2 - \omega y_1 + \frac{\sigma_2}{\sqrt{\varepsilon}} \xi_2 (t) \, . 
  \label{eq:slowOsc}
\end{eqnarray}

\subsection{Homogenization}

The homogenized equation is similar to the one for the additive triad, with an added
constant forcing $C_r$ in the reduced SDE
\begin{eqnarray*}
  \frac{\partial x}{\partial \theta} & = & C_m x + C_r + \sqrt{2 A_0} \xi
  (t)\\
  C_r & = & \frac{B^{(0)}}{\gamma_1 + \gamma_2} \omega \left(
  \frac{\sigma_2^2}{2 \gamma_2} - \frac{\sigma_1^2}{2 \gamma_1} \right) \, .
\end{eqnarray*}

\subsection{Weak coupling limit}

The coupling functions $\psi_x$ and $\psi_y$ are now
\begin{eqnarray*}
  \psi_x (y) & = & B^{(0)} y_1 y_2\\
  \psi_y (x, y) & = & x \left(\begin{array}{c}
    B^{(1)} y_2\\
    B^{(2)} y_1
  \end{array}\right) + \omega \left(\begin{array}{c}
    y_2\\
    - y_1
  \end{array}\right) \, .
\end{eqnarray*}
The correlation function (\ref{eq.triadCorr}) of the coupling to $x$,
determining the correlations of the parametrization noise $\sigma$ is
\begin{eqnarray*}
  \langle \psi_x (y) \psi_x (f^{\tau} (y)) \rangle & = & {B^{(0)}}^2 \langle y_1
  f^{\tau} (y_1) \rangle \langle y_2 f^{\tau} (y_2) \rangle\\
  & = & {B^{(0)}}^2 \exp (- (\gamma_1 + \gamma_2) s) \frac{\sigma_1^2}{2
  \gamma_1}  \frac{\sigma_2^2}{2 \gamma_2} \, .
\end{eqnarray*}
The response function (\ref{eq.triadMem}) of $\psi_x$ to $\psi_y$, determining
the memory kernel of the parametrization, is similar to the one for the
additive triad, with an added exponential function, the integral of which
gives the same constant $C_r$ of the homogenized equations
\begin{eqnarray*}
  R (\tau, x) & = & \langle \psi_y (x, y) \partial_y \psi_x (y (\tau))
  \rangle\\
  & = & \exp (- \gamma \tau) (D_1 x + D_0)\\
  D_1 & = & B^{(0)} \left( B^{(1)} \frac{\sigma_2^2}{2 \gamma_2} + B^{(2)}
  \frac{\sigma_1^2}{2 \gamma_1} \right) = \gamma C_m\\
  D_0 & = & \omega B^{(0)} \left( \frac{\sigma_2^2}{2 \gamma_2} -
  \frac{\sigma_1^2}{2 \gamma_1} \right) = \gamma C_r  \, .
\end{eqnarray*}
Combined, this then results in the following non-Markovian parametrized equations
\begin{eqnarray}
  \frac{d \tilde{x}}{d \tau} & = & \varepsilon \sigma (\tau) + \varepsilon^2
  \int_0^{\infty} \mathrm{d}s R (s, \tilde{x} (\tau - s)) \nonumber\\
  & = & \varepsilon \sigma (\tau) + \varepsilon^2 \int_0^{\infty} \mathrm{d}s \exp (-
  \gamma s) (D_1  \tilde{x} (\tau - s) + D_0) \nonumber\\
  & = & \varepsilon \sigma (\tau) + \varepsilon^2 \int_0^{\infty} \mathrm{d}s \exp (-
  \gamma s) \tilde{x}  (\tau - s) + \varepsilon^2 C_r \, . \label{eq:slowOscMemParam}
\end{eqnarray}

\subsubsection{Markovian parametrization}

The non-Markovian equation (\ref{eq:slowOscMemParam}) can again be
Markovianized by a two-level Ornstein-Uhlenbeck process of the form
\begin{eqnarray}
  \frac{dz_1}{d \tau} & = & \varepsilon C_1 z_2 \nonumber\\
  \frac{dz_2}{d \tau} & = & - \gamma z_2 + \sigma_z \xi (t) + \varepsilon
  (C_2 z_1 + C_3) \, . \label{eq:slowOscMarkovParam}
\end{eqnarray}
The corresponding correlation and memory terms are
\begin{eqnarray}
  C (\tau) & = & C_1^2 e^{- \gamma \tau} \frac{\sigma_z^2}{2 \gamma} \\
  R (\tau) & = & C_1 e^{- \gamma \tau} (C_2 z_1 + C_3) \, .
\end{eqnarray}
We can therefore take
\begin{eqnarray*}
  C_3 & = & D_0 / C_1\\
  & = & \omega \left( \frac{\sigma_2^2}{2 \gamma_2} - \frac{\sigma_1^2}{2
  \gamma_1} \right) \, .
\end{eqnarray*}

In the limit $\varepsilon \rightarrow 0$ in the Markovian parametrization
(\ref{eq:slowOscMarkovParam}) we again recover the homogenized equations.

\subsection{Exit times}\label{sec.oscExit}

When comparing initial ensemble spread and autocorrelation functions for the
slow variable of this system with the weak coupling parametrization and the
homogenized system, the results are similar to those presented for the
additive triad above. Additionally, here we perform a comparison of a rare
event statistic, the first exit time of the slow variable from an interval $[-
1, 1]$ when the slow variable is initialized at $0$.

\begin{table}[!hbt]
\centering
  \begin{tabular}{|l|l|l|l|}
    \hline
    $\varepsilon$ & 0.5 & 0.25 & 0.125\\
    \hline
    homogenization & 0.403 & 0.184 & 0.0982\\
    \hline
    weak coupling & 0.205 & 0.0839 & 0.0589\\
    \hline
  \end{tabular}\begin{tabular}{l}
    
  \end{tabular}
  \caption{The relative error on the mean exit time $| \mathbb{E}_1 (\tau)
  -\mathbb{E}_0 (\tau) | /\mathbb{E}_0 (\tau)$ where $\mathbb{E}_0 (\tau)$
  is the mean exit time from $[- 1, 1]$ of the full triad system and
  $\mathbb{E}_1 (\tau)$ is the mean exit time of the parametrized systems
  with $B^{(0)} = - 0.75$, $B^{(1)} = - 0.25$, $B^{(2)} = 1$, $\omega = 0.25$,
  $\gamma_1 = 1 / \delta$, $\sigma_1 = \sqrt{2 / \delta}$, $\gamma_2 = 1$ and
  $\sigma_2 = \sqrt{2}$ with $\delta = 0.75$.\label{tab.exitSO}}
\end{table}

\begin{table}[!hbt]
\centering
  \begin{tabular}{|l|l|l|l|}
    \hline
    $\varepsilon$ & 0.5 & 0.25 & 0.125\\
    \hline
    homogenization & 0.420 & 0.217 & 0.115\\
    \hline
    weak coupling & 0.232 & 0.0814 & 0.0395\\
    \hline
  \end{tabular}\begin{tabular}{l}
    
  \end{tabular}
  \caption{The relative error on the standard deviation of the exit times $|
  \sigma_1 (\tau) - \sigma_0 (\tau) | / \sigma_0 (\tau)$ where $\sigma_0
  (\tau)$ is the standard deviation of exit times from $[- 1, 1]$ of the full
  triad system and $\sigma_1 (\tau)$ is the standard deviation of exit times
  of the parametrized systems. Parameters are chosen as in Table
  \ref{tab.exitSO}.\label{tab.exitErrorSO}}
\end{table}

The results in Tables \ref{tab.exitSO} and \ref{tab.exitErrorSO} show that the statistics of
exit times are significantly better approximated in the weak coupling
parametrization.

\section{The rapidly oscillating additive triad}

A further generalization of the additive triad (\ref{eq.triad}) is to
introduce an interaction between the $y$ variables on the fast time scale \citep{doacti}. The
dynamical equations for the rapidly oscillating triad are
\begin{eqnarray}
  \frac{d x_{}}{d t} & = & B^{(0)} y_1 y_2 \nonumber \\
  \frac{d y_1}{d t} & = & B^{(1)} y_2 x - \frac{\gamma_1}{\varepsilon} y_1 +
  \frac{\omega}{\varepsilon} y_2 + \frac{\sigma_1}{\sqrt{\varepsilon}} \xi_1
  (t) \nonumber \\
  \frac{d y_2}{d t} & = & B^{(2)} x y_1 - \frac{\gamma_2}{\varepsilon} y_2
  - \frac{\omega}{\varepsilon} y_1 + \frac{\sigma_2}{\sqrt{\varepsilon}} \xi_2
  (t) \, . \label{eq.rapidlyOsc}
\end{eqnarray}
Note the difference in scaling on the oscillatory terms $\omega y_i$
compared to Eq. (\ref{eq:slowOsc}). The invariant measure of the fast system
is a correlated Gaussian measure $\begin{array}{l}
  \rho (y) = \exp (- y^T  (2 R)^{- 1} y) /\mathcal{Z}
\end{array}$ determined by
\begin{eqnarray*}
  \Gamma R + (\Gamma R)^T & = & \Sigma^T \Sigma
\end{eqnarray*}
with
\begin{eqnarray*}
\Gamma = \left(\begin{array}{cc}
  \gamma_1 & - \omega\\
  \omega & \gamma_2
\end{array}\right)
\end{eqnarray*}
and
\begin{eqnarray*}
\Sigma = \left(\begin{array}{cc}
  \sigma_1^{} & 0\\
  0 & \sigma_2
\end{array}\right) \,.
\end{eqnarray*}

Homogenization leads to a solvability condition on the system \ref{eq.rapidlyOsc} that is fulfilled if either $\omega=0$ or $\sigma_1^2 / \gamma_1 = \sigma_2^2/ \gamma_2$. The homogenized equation is now given by
\begin{eqnarray*}
  \dot{x} & = & C_{\omega} x + \sqrt{2 A_{\omega}} \xi (t)
\end{eqnarray*}
with
\begin{eqnarray*}
  C_{\omega} & = & b B^{(1)} R_{22} + 2 (a B^{(1)} + c B^{(2)}) R_{12} + b
  B^{(2)} R_{11}\\
  A_{\omega} & = & B^{(0)} (3 a R_{11} R_{12} + b (R_{11} R_{22} + R_{12}^2) +
  3 c R_{22} R_{12})\\
  b & = & \frac{B^{(0)}}{\left( \frac{\omega^2}{\gamma_1} +
  \frac{\omega^2}{\gamma_2} + \gamma_1 + \gamma_2 \right)}\\
  a & = & (- \omega / 2 \gamma_1) b\\
  c & = & (\omega / 2 \gamma_2) b \, .
\end{eqnarray*}

\subsection{Weak coupling}

The coupling functions of Eq. (\ref{eq.rapidlyOsc}) have the following form
\begin{align}
  \psi_x (y_1, y_2) & = B^{(0)} y_1 y_2\\
  \psi_y (x, y_1, y_2) & = x (B^{(1)} y_2, B^{(2)} y_1)^T \, .
\end{align}

The correlation function $\langle \psi_x (y_1, y_2)
 \psi_x (y_1(t), y_2(t)) \rangle$ appearing in the weak coupling
expansion can again be calculated explicitly. Solutions of the fast
Ornstein-Uhlenbeck system $\dot{y} = - \Gamma y + \Sigma \xi$ can be written
as
\begin{eqnarray*}
  y_i (t) & = & [e^{- \Gamma t} y (0)]_i + \int_0^t \mathrm{d} \tau [e^{- \Gamma (t -
  \tau)} \Sigma \dot{W} (\tau)]_i \, .
\end{eqnarray*}
Inserting this expression into the autocorrelation function gives
\begin{align*}
  \sigma_{\omega} (t) := \langle \psi_x (y_1, y_2)
  \psi_x (y_1(t), y_2(t)) \rangle & =  (B^{(0)})^2  \langle y_1 (0) y_2 (0) y_1 (t)
  y_2 (t) \rangle\\
  & = (B^{(0)})^2  \Bigl( [e^{- \Gamma t}]_{11} [e^{- \Gamma t}]_{21} (3
  R_{11} R_{12}) \Bigr. \\
  & \hspace{3cm} + \left( [e^{- \Gamma t}]_{11} [e^{- \Gamma t}]_{22} + [e^{-
  \Gamma t}]_{12} [e^{- \Gamma t}]_{21} \right) (R_{11} R_{22} + 2 R_{12}^2) \\
  & \hspace{3cm} + \Bigl. [e^{- \Gamma t}]_{12} [e^{- \Gamma t}]_{22} (3 R_{22} R_{12}) \Bigr) \\
  & \hspace{3cm} + (B^{(0)})^2 R_{12} \int_0^t \mathrm{d} \tau_1 d \tau_2  \langle
  [e^{- \Gamma t} \Sigma \xi (\tau_1)]_1 [e^{- \Gamma t} \Sigma \xi
  (\tau_2)]_2 \rangle  \, ,
\end{align*}
since the noise $\xi$ is white and has zero mean.

The memory term $h$ can be calculated by performing integration by parts on
the response function, resulting in a fluctuation-dissipation type expression:
\begin{eqnarray*}
  h_{\omega} (\tau) & = & \left\langle \left( - \frac{\nabla . (\rho
  \psi_y)}{\rho} \right) \psi_x (\tau) \right\rangle\\
  & = & B^{(0)} x \left\langle \left( B^{(1)} [R^{- 1}]_{12} y_2^2 (0) +
  (B^{(1)} [R^{- 1}]_{11} + B^{(2)} [R^{- 1}]_{22}) y_1 (0) y_2 (0) + B^{(2)}
  [R^{- 1}]_{12} y_1^2 (0) \right) y_1 (\tau) y_2 (\tau) \right\rangle
\end{eqnarray*}

\subsubsection{Markovian parametrization}

Guided by the Markovian form of the previous triad systems, we again want to
derive a Markovian parametrization with a reduced one-level Ornstein-Uhlenbeck
system as the fast component:
\begin{eqnarray}
  \dot{z}_1 & = & \varepsilon C_1 z_2 \nonumber\\
  \dot{z}_2 & = & \varepsilon C_2 z_1 - \gamma z_2 + \sigma_z \xi_z (t) \, .
  \label{eq.fastMarkov}
\end{eqnarray}
In this case, there is no exact match between the auto-correlation and
response functions of this Markovian system and the non-Markovian weak
coupling parametrization. The choice of the parametrization parameters is
therefore not exactly determined and one needs to choose a parametrization
such that the auto-correlation and response functions of the coupling function
in the fast component of the full system are approximated in some sense. A
further restriction comes from the fact that in the limit $\varepsilon
\rightarrow 0$ the limiting path distribution of the full system is determined
by the homogenized equation and we therefore want to retain this limiting
behavior in the parametrized system. To have this limiting property, we have
the following constraints on the parameters in Eq. (\ref{eq.fastMarkov})
\begin{eqnarray*}
  \frac{C_1^2 \sigma_z^2}{2 \gamma^2} & = & A_\omega\\
  \frac{C_1 C_2}{\gamma} & = & C_\omega \, ,
\end{eqnarray*}
where $A_\omega$ and $C_\omega$ are the forcing and friction parameters obtained through
homogenization. For formal equivalence between the reduced and full equations,
we furthermore set $C_1 = B^{(0)}$. With the remaining free parameters we can
match the response and correlation functions in a more precise manner, for
example by matching the values of these functions at time $t = 0$. In this way,
we get
\[ C_2 = \frac{h_y (0)}{B^{(0)}} \]
and
\[ \sigma_z^2 = \frac{2 \gamma_{\text{wc}} \sigma_{\omega} (0)}{B^{(0) 2}} \, , \]
where $h_y = h_{\omega} / x$.

\begin{figure}[t]
\includegraphics[width=8.3cm]{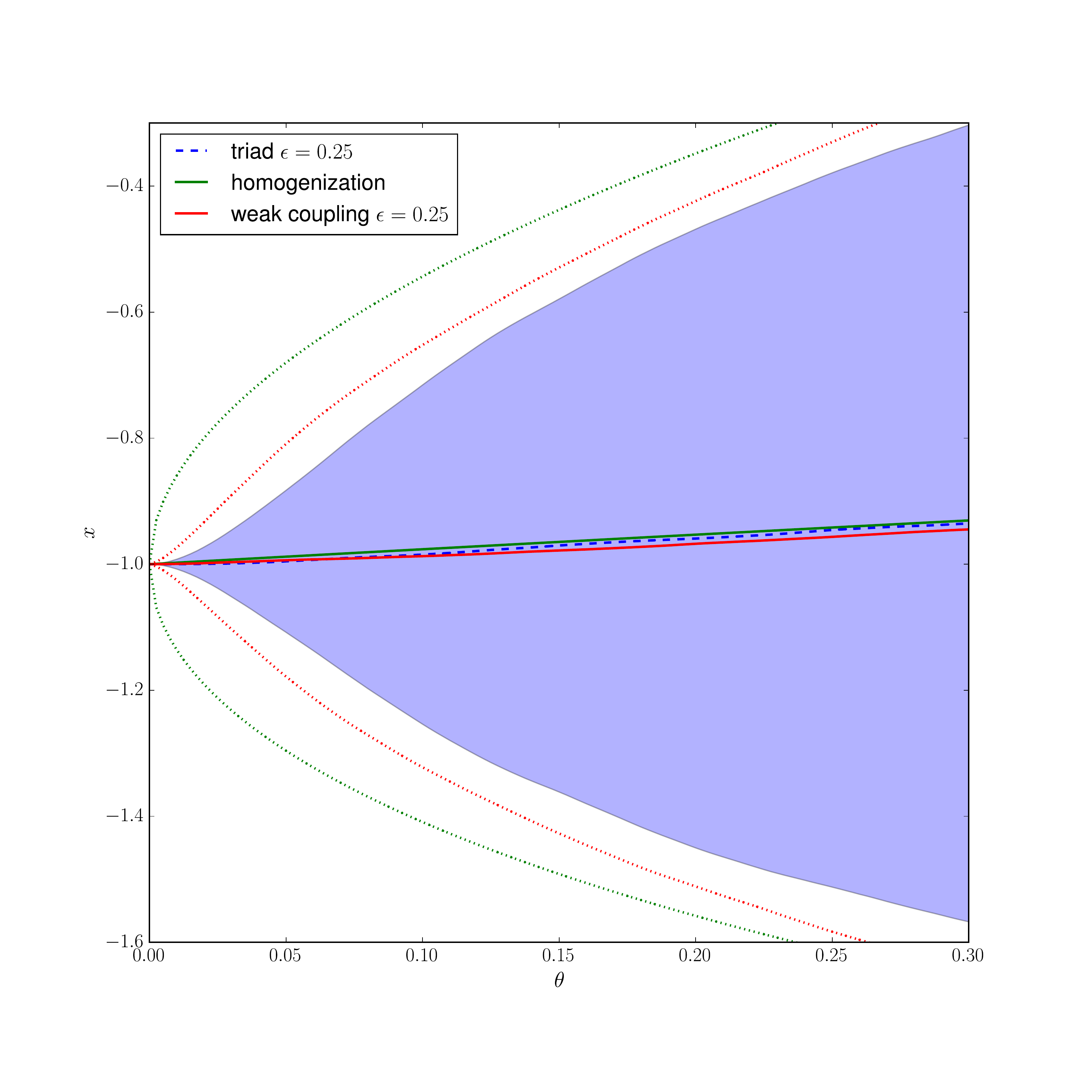}
\caption{comparison of the ensemble spread for the original oscillating
  triad system for $\varepsilon = 0.25$ from an initial condition (-5,0,0)
  (the ensemble mean is the blue dashed line, $2 \sigma$ interval the blue
  shaded area), the two-level Ornstein-Uhlenbeck process from the weak
  coupling method (\ref{eq.memoryparamMarkov}) from an initial condition
  (-5,0) (ensemble mean: red dash-dotted line, $2 \sigma$ interval: red
  dash-dot-dotted lines) and the one-level Ornstein-Uhlenbeck process from
  homogenization (\ref{eq.fastMarkov}) from $x = - 5$ (ensemble mean: green
  solid line, $2 \sigma$ interval: dotted lines) $B^{(0)} = - 0.75$, $B^{(1)}
  = - 0.25$, $B^{(2)} = 1$, $\omega = 1 / 12$, $\gamma_1 = 1 / \delta$,
  $\sigma_1 = \sqrt{2 / \delta}$, $\gamma_2 = 1$ and $\sigma_2 = \sqrt{2}$
  with $\delta = 0.75$.\label{fig.fastEnsemble}}
\end{figure}

A simulation of the ensemble spread from a fixed initial condition is shown in
Figure \ref{fig.fastEnsemble}. It demonstrates that the weak coupling
parametrization (\ref{eq.fastMarkov}) outperforms the homogenized reduced
system.

\subsection{Exit times}

The same experiment on exits from an interval has been performed as described in Section
\ref{sec.oscExit}. The results are displayed in Table \ref{tab.exitFO}.
As before, the weak coupling reduced system gives a much better result when
compared to the homogenized system.

\begin{table}[!hbt]
  \centering
  \begin{tabular}{|l|l|l|l|}
    \hline
    $\varepsilon$ & 0.5 & 0.25 & 0.125\\
    \hline
    homogenization & 0.534 & 0.262 & 0.118\\
    \hline
    weak coupling & 0.322 & 0.127 & 0.0619\\
    \hline
  \end{tabular}\begin{tabular}{l}
    
  \end{tabular}
  \caption{The relative error on the mean exit time $| \mathbb{E}_1 (\tau)
  -\mathbb{E}_0 (\tau) | /\mathbb{E}_0 (\tau)$ where $\mathbb{E}_0 (\tau)$
  is the mean exit time from $[- 1, 1]$ of the full triad system and
  $\mathbb{E}_1 (\tau)$ is the mean exit time of the parametrized systems.
  The parameters are the same as those used for Fig.
  \ref{fig.fastEnsemble}.\label{tab.exitFO}}
\end{table}

\begin{table}[!hbt]
  \centering
  \begin{tabular}{|l|l|l|l|}
    \hline
    $\varepsilon$ & 0.5 & 0.25 & 0.125\\
    \hline
    homogenization & 0.583 & 0.286 & 0.118\\
    \hline
    weak coupling & 0.362 & 0.109 & 0.0503\\
    \hline
  \end{tabular}\begin{tabular}{l}
    
  \end{tabular}
  \caption{The relative error on the standard deviation of the exit times $|
  \sigma_1 (\tau) - \sigma_0 (\tau) | / \sigma_0 (\tau)$ where $\sigma_0
  (\tau)$ is the standard deviation of exit times from $[- 1, 1]$ of the full
  triad system and $\sigma_1 (\tau)$ is the standard deviation of exit times
  of the parametrized systems. The parameters are the same as those used for
  Fig. \ref{fig.fastEnsemble}.\label{tab.exitErrorFO}}
\end{table}

\section{The multiplicative triad}

A final type of interactions is given by the multiplicative triad
equations \citep{MTV02}
\begin{eqnarray}
  \frac{d x_1}{d t} & = & B^{(1)} x_2 y  \nonumber\\
  \frac{d x_2}{d t} & = & B^{(2)} x_1 y \nonumber\\
  \frac{d y}{d t} & = & B^{(3)} x_1x_2 - \frac{\gamma_m}{\varepsilon} y +
  \frac{\sigma_m}{\sqrt{\varepsilon}} \xi (t) \, ,  \label{eq.mtriad}
\end{eqnarray}
which describes the interplay between two $x$ modes and a
stochastically forced single $y$ mode. In the absence of forcing and
dissipation energy conservation is satisfied if $\sum_i B^{(i)}=0$.
In the system (\ref{eq.mtriad}) the $y$ mode can be eliminated
directly by integrating the last equation of (\ref{eq.mtriad})
\begin{eqnarray}
 y(t) = e^{-\frac{\gamma_m}{\varepsilon} t} y(0) 
+ \int_0^t \mathrm{d} t' \left(\frac{\sigma_m}{\sqrt{\varepsilon}} \xi(t') +
 B^{(3)} x_1(t') x_2(t')  \right) e^{-\frac{\gamma_m}{\varepsilon} (t -t')}\, . \nonumber
\end{eqnarray}
Inserting this result in the equations for the $x$ variables, one obtains
\begin{eqnarray}
\label{analytic_mtriad}
\frac{d}{dt}
\left(\begin{array}{c}
    x_1(t)\\
    x_2(t)
  \end{array}\right)
&=  
\left(\begin{array}{c}
    B^{(1)} x_2(t)\\
    B^{(2)} x_1(t)
  \end{array}\right)
\left \{
e^{-\frac{\gamma_m}{\varepsilon} t} y(0) 
+ \int_0^t \mathrm{d} t' \left(\frac{\sigma_m}{\sqrt{\varepsilon}} \xi(t-t') +
 B^{(3)} x_1(t - t') x_2(t - t')  \right) 
e^{-\frac{\gamma_m}{\varepsilon} t'}
\right \}\, .
\end{eqnarray}
Note that the first two term on the righthand side. result from a Ornstein-Uhlenbeck process with zero mean and stationary time autocorrelation function given by
$
\frac{\sigma_m^2}{2 \gamma_m}e^{-\frac{\gamma_m}{\varepsilon} t}.
$

\subsection{Weak coupling}
The coupling functions for the multiplicative triad read
\begin{eqnarray}
\psi_x(x, y) &= (B^{(1)} x_2 y, B^{(2)} x_1 y)^T\, ,\nonumber\\
\psi_y(x)  &= B^{(3)} x_1 x_2\, .\nonumber
\end{eqnarray}
The coupling terms in the $x$ equations are separable
\begin{align}
\psi_{x,i}(x,y) = a_i \psi'_{x,1,i}(x) \psi'_{x,2,i}(y)
\end{align}
with $\langle \psi'_{x,2,i}(y) \rangle_{\rho_{OU}} = 0$, where
\begin{align}
  a_1 &= B^{(1)}\, ,\quad 
\psi'_{x,1,1}(x) = x_2\, ,\quad 
\psi'_{x,2,1}(y) = y\, ,\nonumber\\
  a_2 &= B^{(2)}\, ,\quad 
\psi'_{x,1,2}(x) = x_1\, ,\quad 
\psi'_{x,2,2}(y) = y\, .\nonumber
\end{align}
The resulting parametrization in the weak coupling approach {\citep{wouters_disentangling_2012,wouters_multilevel_2013}} reads
\begin{eqnarray}
  \frac{d x_i}{d \tau} & = & \varepsilon a_i \psi'_{x,1,i} \sigma_i (\tau)
 + \varepsilon^2 \int_0^{\infty} \mathrm{d} s R_i (s, x (\tau - s)) \, ,  \label{eq.mult_memoryparam}
\end{eqnarray}
with a noise term $\sigma_i$ with zero mean and correlation given by
\begin{align}
\langle \sigma_i(0) \sigma_j(\tau)\rangle  = 
\langle \psi'_{x,2,i}(y) \psi'_{x,2,j}(y(\tau)) \rangle_{\rho_{OU}}
= \frac{\sigma_m^2}{2 \gamma_m} e^{- \gamma_m \tau}\, . \nonumber
\end{align}
The memory kernel has the form
\begin{align}
  R_i(s, x) = \langle \psi_y(x,y) \cdot 
\nabla_y \psi_{x,i} (x(s), y(s)) \rangle_{\rho_{OU}}\, , \nonumber\\
R(s, x) = B^{(3)} x_1 x_2 e^{-\gamma s}
\left(\begin{array}{c}
    B^{(1)} x_2(s)\\
    B^{(2)} x_1(s)
  \end{array}\right) \nonumber
\end{align}
Thus (\ref{eq.mult_memoryparam}) can be written as
\begin{eqnarray}
\label{eq.weak_coupling_mtriad}
\frac{d}{d\tau}
\left(\begin{array}{c}
    x_1(\tau)\\
    x_2(\tau)
  \end{array}\right)
&=  
\left(\begin{array}{c}
    B^{(1)} x_2(\tau)\\
    B^{(2)} x_1(\tau)
  \end{array}\right)
\left \{
\sigma(\tau)
+ \int_0^\infty \mathrm{d} s
 B^{(3)} x_1(\tau - s) x_2(\tau - s) e^{-\gamma_m s}
\right \}\, ,
\end{eqnarray}
which is exactly the same result as in (\ref{analytic_mtriad}), if we
rescale time and assume as initial condition $x_1(t)=x_2(t)=0$
for $t<0$. In this case the weak coupling approach recovers exactly the full model. The original three component system was reduced to a two component non-Markovian system but there is no efficiency gain using the parametrization since the corresponding Markovian system is again a three component one.

\subsection{Homogenization} 
Introducing a longer time scale $\theta = \varepsilon^{2} \tau$ in 
(\ref{eq.weak_coupling_mtriad}) and taking the limit $\varepsilon \to 0$ one recovers the homogenization result in Stratonivich formulation 
\begin{align}
\frac{d}{d\theta}
\left(\begin{array}{c}
    x_1\\
    x_2
  \end{array}\right)   =  \frac{B^{(3)}}{\gamma} x_1 x_2 
 \left(
 \begin{array}[h]{c}
   B^{(1)} x_2 \\
   B^{(2)} x_1 
 \end{array}
 \right) 
+ \frac{\sigma_m}{\gamma_m}
 \left(
 \begin{array}[h]{c}
   B^{(1)} x_2 \\
   B^{(2)} x_1 
 \end{array}
 \right) \xi(\theta) \, .
\end{align}
The latter corresponds to an It{\^o} stochastic differential equation
of the form
\begin{align}
\frac{d}{d\theta}
\left(\begin{array}{c}
    x_1\\
    x_2
  \end{array}\right)   =  \frac{B^{(3)}}{\gamma} x_1 x_2 
 \left(
 \begin{array}[h]{c}
   B^{(1)} x_2 \\
   B^{(2)} x_1 
 \end{array}
 \right) 
+ \frac{\sigma_m^2}{2 \gamma_m^2}
 B^{(1)} B^{(2)} \left(
 \begin{array}[h]{c}
   x_1 \\
   x_2 
 \end{array}
 \right)
+ \frac{\sigma_m}{\gamma_m}
 \left(
 \begin{array}[h]{c}
   B^{(1)} x_2 \\
   B^{(2)} x_1 
 \end{array}
 \right) \xi(\theta)\, .
\end{align}
For a comparison of the statistics of the multiplicative triad and of the homogenization model we refer to \citep{MTV02}.

\conclusions

In this work we have worked out a first application of the weak coupling
response method of {\citep{wouters_disentangling_2012,wouters_multilevel_2013}}
to a multiscale stochastic system. By the choice of system we were able to
perform both homogenization and the weak coupling reduction on this system,
thereby allowing for direct comparison between the two reductions.

The response method yields a non-Markovian equation, making it cumbersome to
integrate numerically. We have demonstrated here that for the systems studied
the non-Markovian equation can be further reduced to a Markovian equation.
Even with this further reduction the system gives a better match to the
original system than the homogenized equations.

In the case of the additive triads, the system that is obtained through the Markovianization procedure is of
intermediate complexity, between the full system and the homogenized limit. In
the systems studied here, the retention of a fast time scale in the reduced
system means that the reduction in simulation complexity is modest (one
variable instead of two and a linear coupling instead of a nonlinear one). In the case of the multiplicative triad the weak coupling parametrization recovers exactly the full model and there is no efficiency gain. In
many applications of practical relevance, however, one considers situations
where the number of degrees of freedom of the unresolved variables is
considerably larger than those of the slow variables of interest. A reduction
to a system of one or a few variables will constitute a significant reduction
in complexity in this case. This approach can be compared to the
superparametrization approach to convection, where convection is parametrized
by a model that is still dynamical in nature, yet significantly simpler than
the full convective motion
{\citep{randall_breaking_2003,grooms_efficient_2013,grooms_stochastic_2014}}.

\begin{acknowledgements}
JW is grateful to Georg Gottwald and Cesare Nardini for stimulating
discussions. VL is grateful to M Chekroun, C Franzke, and M Ghil for a lot of
food for thought on the problem of constructing reduced model in geophysical
fluid dynamical systems.

The research leading to these results has received funding from the European
Community's Seventh Framework Programme (FP7/2007-2013) under grant agreement
n{\textdegree} PIOF-GA-2013-626210, as well as from the DFG project MERCI. SD if thankful to the German Research Foundation (DFG) for partial support through DO 1819/1-1. UA thanks the
German Research Foundation (DFG) for partial support through
grant AC 71/7-1.
\end{acknowledgements}

\bibliographystyle{copernicus}
\bibliography{triad}

\end{document}